\documentclass[aps,prb,preprint,superscriptaddress,showpacs]{revtex4-1}
 \usepackage{graphicx,psfrag}
 \usepackage[latin1]{inputenc}
 \usepackage{natbib}
 \usepackage{amsmath, amsthm, amssymb}
 \usepackage{color}
 \usepackage{comment}
 \usepackage{hyperref}
 
 \usepackage{setspace}

 \newcommand{\ds}{\displaystyle}
 
 \newcommand{\notes}[1]{}

 \newcommand{\beq}{\begin{equation}}
 \newcommand{\eeq}{\end{equation}}
 \newcommand{\beqnn}{\begin{equation*}}
 \newcommand{\eeqnn}{\end{equation*}}
 \newcommand{\beqas}{\begin{eqnarray*}}
 	\newcommand{\eeqas}{\end{eqnarray*}}
 \newcommand{\beqa}{\begin{eqnarray}}
 \newcommand{\eeqa}{\end{eqnarray}}

\begin{document}

\title{Generation and stability of dynamical skyrmions and droplet solitons}

\author{Nahuel Statuto}
\affiliation{Institut de Ci\`encia de Materials de Barcelona (ICMAB-CSIC), Campus UAB, 08193 Bellaterra, Spain}
\affiliation{Dept. of Condensed Matter Physics, University of Barcelona, 08028 Barcelona, Spain}
\author{Joan Manel Hern\`andez}
\affiliation{Dept. of Condensed Matter Physics, University of Barcelona, 08028 Barcelona, Spain}
\author{Andrew D. Kent}
\affiliation{Center for Quantum Phenomena, Department of Physics, New York University, New York, New York 10003 USA}
\author{ Ferran Maci\`a}
\email{fmacia@icmab.es}  
\affiliation{Institut de Ci\`encia de Materials de Barcelona (ICMAB-CSIC), Campus UAB, 08193 Bellaterra, Spain}

\date{\today}

\begin{abstract}

A spin-polarized current in a nanocontact to a magnetic film can create collective magnetic oscillations by compensating the magnetic damping. In particular, in materials with uniaxial magnetic anisotropy, droplet solitons have been observed a self-localized excitation consisting of partially reversed magnetization that precesses coherently in the nanocontact region. It is also possible to generate topological droplet solitons, known as \emph{dynamical skyrmions}. Here we study the conditions that promote either droplet or dynamical skyrmion formation and describe their stability in magnetic films without Dzyaloshinskii-Moriya interactions. We show that Oersted fields from the applied current as well as the initial magnetization state can determine whether a droplet or dynamical skyrmion forms. Dynamical skyrmions are found to be more stable than droplets. We also discuss electrical characteristics that can be used distinguish these magnetic objects.

\end{abstract}

\maketitle

The control of magnetic states in nanostructures without using magnetic fields is now possible with the discovery of the spin-transfer torque (STT) effect \cite{Slonczewski1996,Berger1996,Katine2000}. A spin-polarized current can transfer angular momentum to a magnetic material \cite{Ralph2008} and modify its magnetization. The STT effect is used in the control of both static and dynamic magnetic states; one can switch the magnetization direction of a magnetic layer within a  nanopillar or create coherent spin waves in an extended thin film \cite{Brataas2012}. In particular the STT effect can be used to nucleate and control solitonic modes---magnetization states that behave as particles. These self-localized magnetic objects include magnetic domains, vortices, bubbles, or skyrmions 
\cite{Myers867,Shibata_prb_2006,Klaui_apl_2006,Khvalkovskiy_prb_2013,Sinova_jap2016,Emori_natureMat2013,Ryui_natureNano2013,perez_apl_2014} and they are receiving a growing interest since they can be topological and, thus, more stable against perturbations such as thermal fluctuations or fabrication defects \cite{Roszler_nature_2006,Nagasoa,Sampaio2013}.  Besides the possibility of nucleation and control of static solitonic modes, the STT effect is also used to excite their dynamical counterparts consisting in oscillating modes that are unstable in dissipative materials---damping is present in all magnetic materials and suppresses these excitations. However, damping can be now compensated locally by the STT effect, for example, with an electrical point contact providing a spin-polarized current \cite{Slavin_prl_2005,Demidov2012,Backes2015,Bonetti2015}.

Dissipative magnetic droplet solitons (droplets hereafter) are nonlinear localized wave excitations consisting of partially reversed precessing spins that can be created in films with perpendicular magnetic anisotropy (PMA) \cite{Hoefer2010}. Droplets have been experimentally created using the STT effect in electric nanocontacts to PMA films \cite{Mohseni2013, Chung2014, Mohseni2014, Macia2014, Lendinez2015, Chung2015, Akerman2016NatComm}. Droplets are magnetic nano-oscillators and have a growing interest as key elements in neuromorphic computation \cite{GrollierIEEE2016,JulieNature2017} and in communication devices \cite{Maiden2014}. Droplets are topologically trivial objects---they can be created continuously from a uniform ferromagnetic state where all spins are aligned in the same direction. A similar magnetic object having topologically non-trivial spin texture could be created in a similar experimental geometry: a \textit{dynamical skyrmion} (DS). Zhou \textit{et al.} \cite{Zhou_natComm_2015} have shown with micromagnetic simulations that DS can be nucleated and sustained with a spin-polarized current in a nanocontact and are, indeed, fundamental solutions for the magnetization excitations in a film with PMA \cite{Ivanov1976}. Liu \textit{et al.} \cite{Sergei2015} presented an experimental observation of a solitonic mode modulation that could indicate the existence of a DS. So far, the topology modification of droplets has been associated to the Dzyaloshinskii-Moriya interaction (DMI) present in some magnetic films \cite{pulia2015}.

\begin{figure}[ht!]
  \centering
  \includegraphics[width=0.65 \columnwidth]{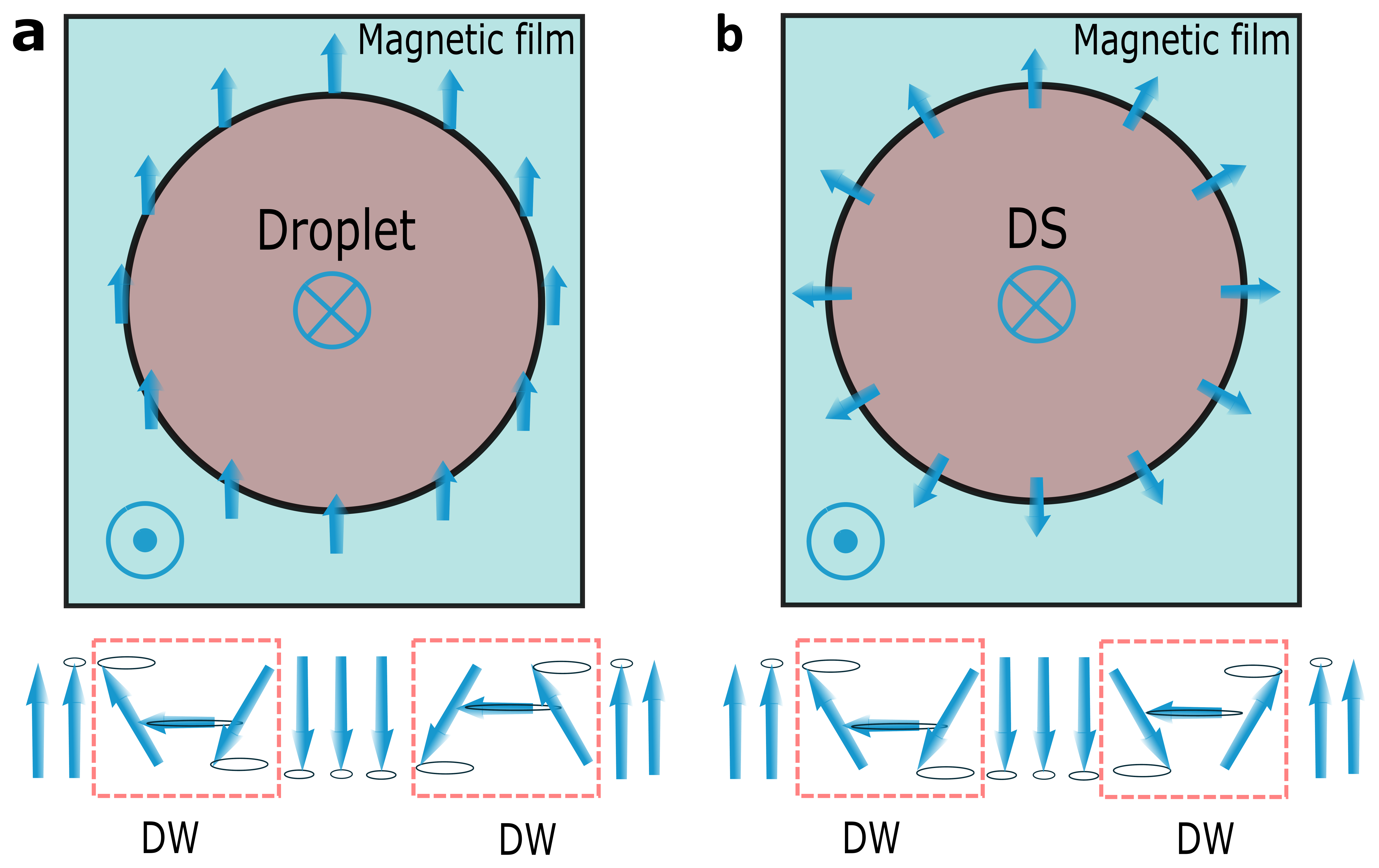}
     \caption{\textbf{Schematic representation of droplet (a) and dynamical skyrmion (DS) (b) magnetic configuration}. Magnetization within droplet or DS is reversed with respect to the film's magnetization and is precessing with a small amplitude at the center and with a larger amplitude at the boundaries. Lower panels show a transversal cut of the spin configuration for the droplet ($S=0$) in a) and DS ($S=1$) in b). }
     \label{fig1}
\end{figure}

A schematic plot of both solitonic modes, droplet and DS, is shown in Fig.\ \ref{fig1} where the blue region represents magnetization pointing out-of-plane and the brown, in the opposite direction. The magnetization of a droplet or a DS is precessing, with a small amplitude near its center and with a larger amplitude at the boundaries. The lower panels of Figs.\ \ref{fig1}a and \ref{fig1}b show the magnetization orientation in a transversal cut of both droplet and DS. The main difference is in the region separating the center of soliton from the rest of film's magnetization; droplets have no topology (the magnetization shown in Fig.\ \ref{fig1}a can be transformed continuously into a ferromagnetic state with all spins aligned in any arbitrary direction) whereas the DS has topology (such a transformation is not possible). The topology in two dimensions can be described by the skyrmion number ($S$), which is calculated mathematically as $S=\ds -\frac{1}{4\pi}\int \mathbf{m}\Big(\partial_x \mathbf{m}\times\partial_y \mathbf{m}\Big)$. A droplet has $S=0$, and a DS has $S=1$.

Here, we investigate the conditions that lead to either droplet or DS formation and we study their stability in nanocontacts to ferromagnetic thin films with PMA and without interfacial DMI. Our micromagnetic simulations show that the Oersted fields associated with the localized electrical current, the initial magnetization state, and the rise time of the injected current, play a key role on determining whether droplet or DS form. DS are more stable to perturbations and can be sustained with much lower currents than droplets. We also provide characteristic features of droplets and DS that could distinguish the two magnetic objects experimentally.

\section*{Results}

\subsection*{Simulations details}

We consider a circular nanocontact to a ferromagnetic thin film with PMA. The parameters for the material are taken from experiments using Co and Ni multilayers \cite{Macia2012,Macia2014}. Magnetization saturation, $M_s=5\times 10^5$ A/m, damping constant, $\alpha=0.03$, uniaxial anisotropy constant, $K_u=2\times 10^5$ J/m$^3$, exchange stiffness constant, $A=10^{-12}$ J/m, and a nanocontact diameter of $150$ nm for most of the presented results. We modeled the magnetization dynamics in the nanocontact by solving the Landau-Lifshitz equation adding the STT term \cite{Slonczewski1996} with a constant spin polarization. We performed micromagnetic simulations using the open-source MuMax code \cite{Vansteenkiste-mumax} using a graphics card with 2048 processing cores. We considered the effects of Oersted fields but we did not include interfacial DMI or temperature effects (full codes are available in Supplementary Materials).

\subsection*{Creation Process}

To excite a droplet in a ferromagnetic layer with PMA using a spin polarized current in a nanocontact, the spin-transfer torque must compensate the damping. There is a threshold current that depends on the NC size, the magnetization, the spin polarization of the current, and the external field \cite{Hoefer2010,Lendinez2015,Akerman2016NatComm,Lendinez2017}. For currents above the threshold, the magnetization in the NC forms a droplet state in a process that can take less than a nanosecond \cite{Jinting_2017}. Once the droplet is created, the current in the nanocontact is still required to sustain the magnetic excitation---although smaller current values than the threshold current are needed \cite{Mohseni2013,Macia2014}. Droplet states can be inferred by measuring the dc resistance of the nanocontact---a reversal of the magnetization produces a change in the nanocontact resistance \cite{Mohseni2013, Chung2014, Mohseni2014, Macia2014, Lendinez2015, Chung2015, Akerman2016NatComm}. Further, the magnetization dynamics of droplets can be detected experimentally through the ac electrical resistance oscillations in the nanocontact \cite{Mohseni2013, Chung2014, Mohseni2014, Macia2014, Chung2015, Akerman2016NatComm} caused by the precessing magnetization in the droplet.

\begin{figure}[htb!]
	\centering
	\includegraphics[width=0.8 \columnwidth]{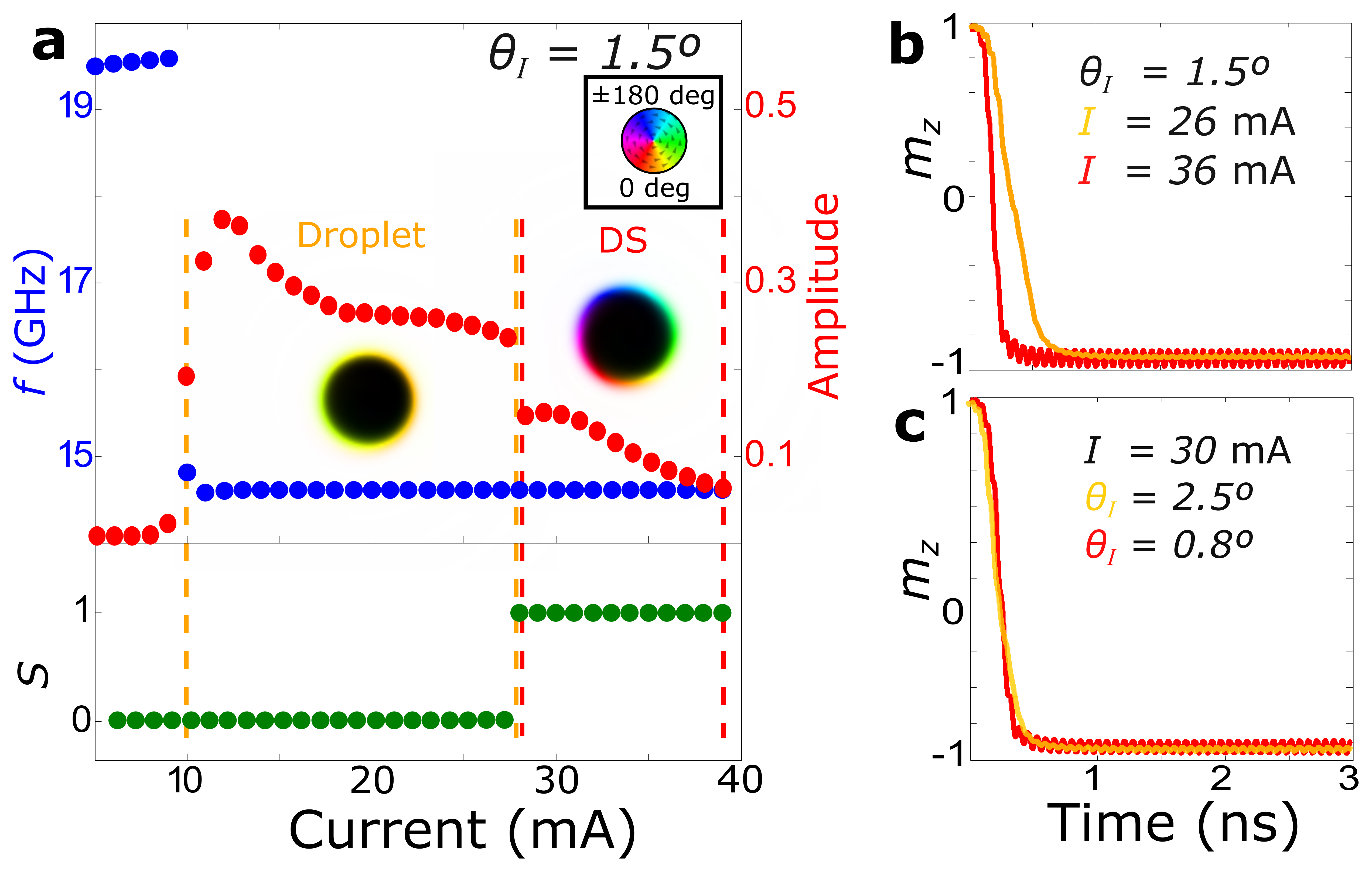}
	\caption{ \textbf{Droplet and dynamical skyrmion creation process}. a) Resonance frequency (in blue dots) and amplitude (red dots) as a function of the applied current for the nanocontact overall magnetization. Both frequency and amplitude correspond to the average over the nanocontact of one of the in-plane components of the magnetization, $m_{x,y}$. The current values are always applied from a same initial magnetization angle in an applied field of 0.5 T and with a polarization of $p=0.45$. At current values below the threshold (below 10 mA) the nanocontact magnetization precesses close to the ferromagnetic resonance frequency with a small amplitude. A first current threshold at 10 mA corresponds to a droplet formation and shows a much larger amplitude (red curve) and a frequency jump down to a lower value---that remains almost constant with increasing the applied current. A second current threshold at 28 mA corresponds to the DS formation and has a similar precession frequency and a smaller amplitude. The bottom panel show the skyrmion number, $S$, at each current step. b) and c) Time evolution of the normalized magnetization inside the NC for droplet (yellow line) and DS (red line) for the same external applied field of $0.5$ T. In b) both solitons are excited at an initial magnetization angle, $\theta_I=1.5^\circ$ but using different applied currents.	In c) both solitons are exited at $30$ mA but changing the initial magnetization angle.}
	\label{fig2}
\end{figure}

DS may also form in a NC to a ferromagnetic layer with PMA \cite{Zhou_natComm_2015,Sergei2015} when a sufficiently large current is applied. The difference between droplet and DS is in the topology of the spins on the boundary that might provide additional stability. For this reason, we are interested in determining the differences in stability between droplet and DS as well as the experimental conditions under with DS form.

In simulations we choose an initial magnetization that is close to equilibrium---all spins aligned with the applied field---and then we apply a spin-polarized current and record the evolution of the magnetization in an area that is 5 times larger than the contact diameter. Figure\ \ref{fig2}a shows the magnetization precession frequency within the NC as a function of the applied spin-polarized current under an applied field of 0.5 T.

For values of current below 10 mA the magnetization in the NC (the average value) has a small oscillation with a frequency close to the ferromagnetic resonance frequency. Above 10 mA there is an abrupt decrease of the frequency together with an increase of the precession amplitude, which corresponds to the creation of a droplet. If we continue applying current of larger amplitude (always starting from a same initial state), we reach a second threshold at 28 mA where DS form, having an almost identical magnetization precession frequency (blue dots) but a much smaller precession amplitude (red dots). The precession amplitude of spins is much larger at the boundary of the soliton than in the central part. Thus, the average nanocontact precession amplitude is mostly driven by the edge precession. In DS the spins at the boundary precess at a similar amplitude than in droplets but the fact that they are not in phase causes a cancellation of the effect when measuring the average contact electrical characteristics---which is a feature to identify DS experimentally. The same arguments applies to describe the smooth decrease in the precession amplitude of the NC magnetization in the droplet state as the current increases from 10 mA to 28 mA; the phase of droplet becomes less and less uniform along the overall droplet edge when increasing the applied polarized current \cite{Jinting_2017}.

The second threshold was also identified by Zhou \textit{et al.} \cite{Zhou_natComm_2015} through mciromagnetic simulations where a DS was excited from an initial ferromagnetic state obtained after a relaxation process. A relaxation process leads to a state with magnetization almost perpendicular to the film with an angle $\theta_I \approx 0^\circ$. In our simulations we indeed study the effect of initial magnetization states on the formation of droplet and DS. The initial magnetization angle, $\theta_I$, is fixed and treated as a parameter in simulations. Figure\ \ref{fig2}a is done using an initial state with $\theta_{I}=1.5^\circ$.

We next study the time evolution of magnetization during the process of droplet and DS formation. Figure\ \ref{fig2}b shows the magnetization evolution in the NC region, $m_z$, as a response of an applied current for a droplet (yellow line) at $26$ mA and for the DS (red line) at $36$ mA; both time traces correspond to points in Fig.\ \ref{fig2}a having an initial magnetization state with $\theta_I=1.5^\circ$. We see that the higher applied current has a faster magnetization reversal, which is something that occurs no matter whether the final state is a droplet or a DS and is caused by a larger STT effect--which is proportional to the applied current \cite{Jinting_2017}. We can also observe that the DS (red line) presents a larger oscillation of the magnetization indicating there is a breathing of the localized object  at the precession frequency \cite{Zhou_natComm_2015,Sergei2015}. We note here that the magnetization $m_z$ average over the NC (plotted in Fig.\ref{fig2}b) is a relevant quantity for experiments as it can be directly associated to the NC resistance.

The initial state determines whether the response to an applied current is a droplet or a DS. In Fig.\ \ref{fig2}c we plot time traces for the magnetization, $m_z$, in the NC for a same applied current, 30 mA, but different initial magnetization states, $\theta_I=2.5^\circ$ (yellow line) and $\theta_I=0.8^\circ$ (red line). We note that the initial state with $\theta_I=2.5^\circ$ evolves to a droplet state whereas the initial state with $\theta_I=0.8^\circ$ evolves to a DS state. The current threshold for DS formation thus has a dependence on the magnetization initial state. Additionally, we measured the precession frequency of droplet and DS for the case presented in Fig.\ \ref{fig2}c. For the same current both the droplet and DC have nearly the same precession frequency: $f=14.60$ GHz for droplet and $f=14.58$ GHz for DS, which is not seen in the transition at 27 mA of Fig.\  \ref{fig2}a due to the small difference. We attribute such a small variation in frequency to the small changes in the size of the magnetic object and therefore in the value of internal magnetic fields---mainly dipolar fields.

\begin{figure}[h!]
  \centering
  \includegraphics[width=0.85\columnwidth]{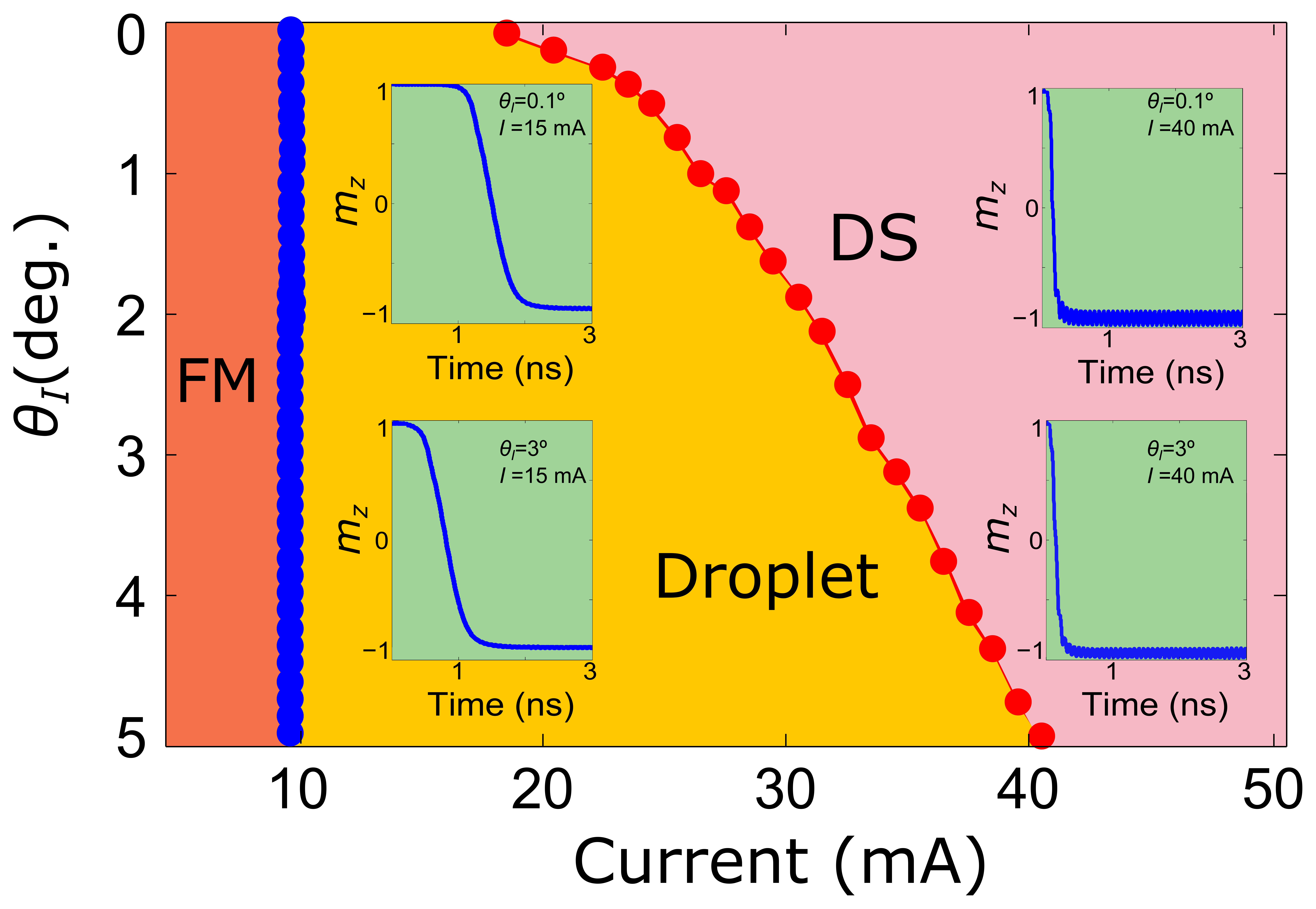}
     \caption{\textbf{Phase diagram of the droplet and DS formation} Creation of both solitonic modes as a function of the applied current, $I$ (with polarization $p=0.45$), and the initial magnetization angle, $\theta_I$. For currents below $10$ mA neither droplet nor DS can be excited, orange region. When the current is higher than $10$ mA a droplet is excited and droplet's threshold current does not depend on $\theta_I$, yellow region. If the current is further increased, DS are created, pink region. The current threshold for DS (red line) is higher than the droplet and depends on the initial magnetization state, $\theta_I$. Insets correspond to time evolution curves of nanocontact magnetization at different conditions.}
     \label{fig3}
\end{figure}

In order to understand how the threshold current for DS formation depends on the initial magnetization state, we repeat the process used in Fig.\ \ref{fig2}a with different initial magnetization states (different of $\theta_I$) and we identify the current that result in a droplet or a DS. Figure\ \ref{fig3}a shows the phase diagram of droplet and DS formation as a function of the applied current and the initial magnetization angle. We see that the threshold for droplet formation is always the same independent of the initial magnetization state; different initial states cause the process of droplet formation to become faster or slower (see traces for time evolution in the insets of Fig.\ \ref{fig3}a) \cite{Jinting_2017}. On the other hand, the threshold for DS formation has a strong dependence on the initial magnetization angle, $\theta_I$, increasing with larger angles. An additional map is provided in the Supplementary materials showing the phase diagram of droplet and DS formation as a function the polarization of the applied current and the initial magnetization angle ($\theta_I$) for a fixed current of $30$ mA. In that case the Oersted-field effects are fixed and only STT effects vary with spin-polarization. At a small polarization, there is a small STT effect and no excitations are present independently of the initial values of magnetization. As the current polarization increases we found first the onset of droplet states and with a further increase the onset of DS. Again the droplet threshold does not depend on the initial state whereas the DS has a strong dependence requiring larger values of polarization at larger angles of the initial magnetization angle, $\theta_I$.

We have used  a polarization of $p=0.45$ for the phase diagram of Fig.\ \ref{fig3} but a different value would shift both droplet and DS thresholds. An increase of polarization from $p=0.45$ to $p=0.6$ produces a shift of 2 mA in the droplet threshold and a shift of 5 mA in the DS threshold. The contact size determines the net current required to excited solitonic modes. We computed droplet and DS thresholds for contact diameters of 50 and 100 nm and obtained values of 4 and 6 mA for the droplet threshold---which represents a decrease of 3 and 5 mA with respect to the diameter of 150 nm presented in Fig.\ \ref{fig3}. Here we note that the thresholds does not scale exactly with the current density beacasue there are always Oersted fields associated with the currents that depend on the contact size. We observed a larger reduction of 5 and 9 mA for the DS formation. Both diagrams are presented in the Supplementary materials.

\subsection*{\label{sec:level1}Stability}

\begin{figure}[h!]
  \centering
  \includegraphics[width=0.85\columnwidth]{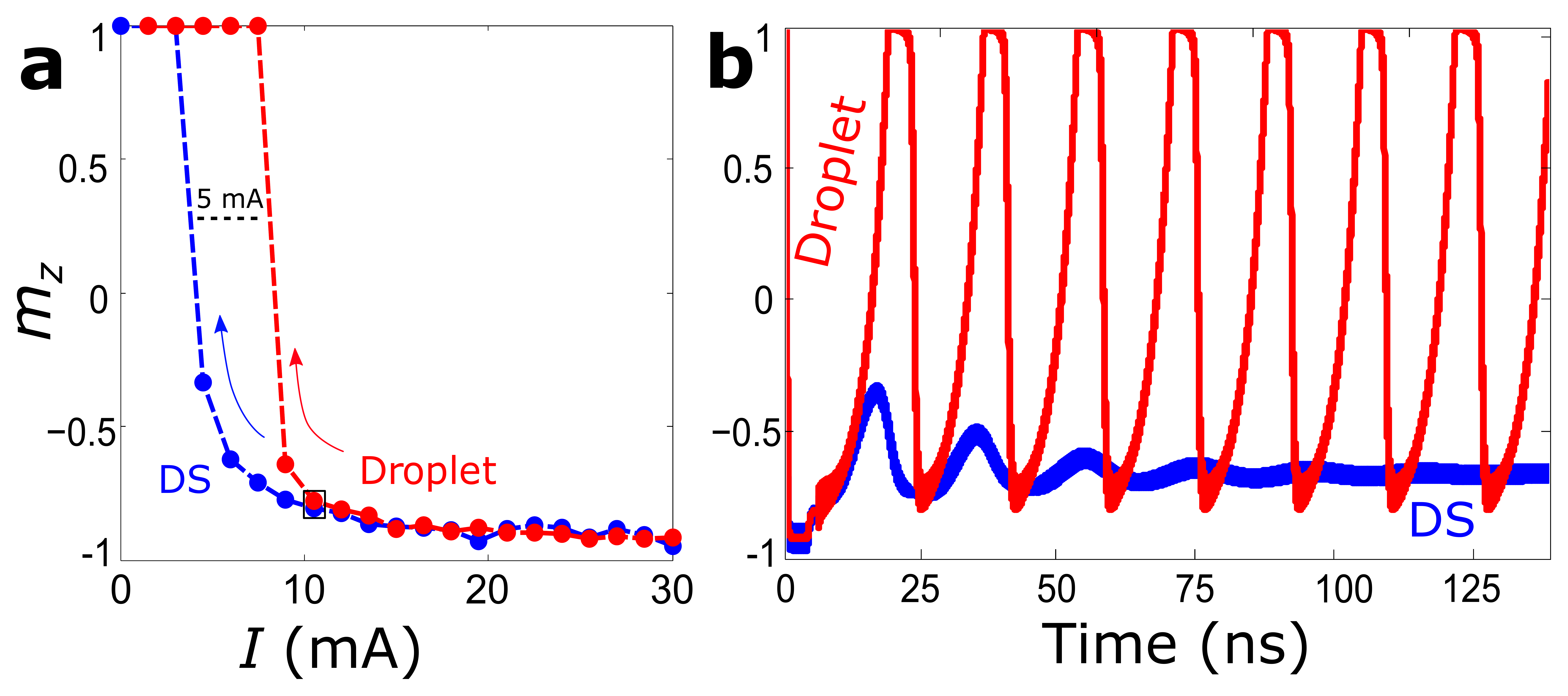}
     \caption{\textbf{Stability of Droplet and DS} a) curves of annihilation of droplet (red line) and DS (blue line) with decreasing the applied current. The curves correspond to the averaged normalized magnetization within the nanocontact, $m_z$, as the applied current decreases from an initial value of $I=30$ mA.  Both droplet and a DS are created at 30 mA (using $\theta_I=3^\circ$ and $\theta_I=0.1^\circ$ respectively). The droplet collapses at about 9 mA whereas the DS does it at a lower current of about 4 mA. b) Time evolution of $m_z$ for a droplet (red line) and a DS (blue line) in the presence of an in-plane magnetic field. Both droplet and DS states are created at $30$ mA using different initial states (same as in a)) and after stabilization the applied current is reduced to $10$ mA, black squares in a), and a small in-plane field of $50$ mT is applied. The magnetization of droplet and DS behaves completely differently; the droplet's magnetization oscillates caused by a drift resonance ($\sim40$ MHz) while the DS's magnetization, although it initially oscillates, it stabilizes after $\sim80$ ns and remains with the initial topology having $S=1$.}
     \label{fig4}
\end{figure}

Both droplet and DS exhibit magnetic bistability over considerable ranges of applied current and magnetic field \cite{Mohseni2013, Chung2014, Mohseni2014, Macia2014, Lendinez2015, Chung2015, Akerman2016NatComm,Zhou_natComm_2015,Sergei2015}. We investigate here the conditions that produce the annihilation of the solitonic modes when a lower degree of spin transfer torque---a lower current---is applied. In Fig.\ \ref{fig4}a we show two curves corresponding to the average magnetization within the NC, $m_z$, as the applied current decreases from an initial value of $I=30$ mA.  A droplet and a DS are created at 30 mA (using $\theta_I=3^\circ$ and $\theta_I=0.1^\circ$ respectively). The droplet collapses at about 9 mA whereas the DS requires a much lower current value of 4 mA to vanish revealing that the DS remains stable over a larger range of applied currents or in other words, the DS requires smaller current values to be sustained.

Next, we investigate the effect of a magnetic field gradient in the NC. A small constant in-plane field, a small change in anisotropy, or a variation in the film's thickness combined with the Oersted fields from the charge current could result in a gradient of effective magnetic field in the NC that could dephase the precession of magnetization in different locations of the NC and eventually annihilate the magnetic excitation. Experiments revealed that the low frequency noise in droplets \cite{Lendinez2015,Akerman2016NatComm,Lendinez2017} is associated with a periodic process of shifting, annihilation, and creation. Simulations showed that an asymmetry of the effective field causes a drift instability resulting in an oscillatory signal of hundreds of MHz---drift resonances. We excite droplet and DS states at $30$ mA using different initial states (same as in Fig.\ \ref{fig4}a) and after a stabilization period we reduce the applied current until $10$ mA, black squares in Fig.\ \ref{fig4}a. We then apply a small in-plane field of $50$ mT in order to destabilize the solitonic modes. The combination of a fixed in-plane field with the Oersted fields creates an in-plane field gradient in the nanocontact. Figure \ref{fig4}b shows the time evolution of the magnetization for a droplet (red line) and a DS (blue line). The small in-plane applied field causes a shift of the droplet away from the NC followed by a re-nucleation of a droplet state. The process of creation and annihilation is repeated at a frequency in the MHz range ($\sim80$ MHz) \cite{Lendinez2015}. The effect of an in-plane field to the DS is different; DS has an initial change as a result of the abrupt change of the magnetic field but later on the DS stabilizes again. Full videos of the evolution of droplet and DS in Fig.\ \ref{fig4}b are available in the Supplementary Materials. 

\section*{\label{sec:level1}Discussion}

 Two main effects are involved in the magnetization dynamics when a spin-polarized current flows through a nanocontact to a magnetic film. On the one hand, a spin-polarized current of the appropriate polarity interacts with the magnetization via the STT effect trying to align the magnetization in the opposite direction of the applied field. The STT effect is proportional to the non-collinear component of the magnetization with respect to the polarization of the current (\emph{i.e.}, if the magnetization is precisely aligned in the direction of the polarized current, say $z$ for the studied case, there is no effect). On the other hand, the electrical current flowing through the nanocontact causes Oersted fields that curl the magnetization. Here we note that in absence of other effects the magnetization of a PMA layer adopts a configuration with $S=1$ in presence of Oersted fields. The skyrmion configuration provides a topological protection in two dimensions, which is valid for variations of the in-plane components of the magnetization.
 
In the creation process of solitonic modes there is a competition between the two mentioned effects. The STT effect increases rapidly as the magnetization tilts from the state perpendicular to the film plane, $\theta_{I}=0^\circ$, and thus if the initial magnetization state is sufficiently far from such a state, the solitonic mode forms without topology resulting in a droplet state, states with $\theta_I\ne0$ can be prepared by increasing the temperature or by applying a short in-plane field pulse. On the other hand if the initial state is closer to $\theta_{I}=0^\circ$ the effect of STT produces a much slower variation of the magnetization and there is a time lapse where the effect of the Oersted fields provides topology to the magnetization in the NC, which eventually results in the formation of a DS. In summary, the farther from equilibrium the initial magnetization state is, the larger it is the required current density to create a DS. There is another ingredient that plays a role in defining whether a droplet or a DS forms; the speed of ramping the polarized current from zero, or from a small value, to a high value that nucleates solitonic modes. Simulations in the diagram shown in Fig.\ \ref{fig2}b and c are done with a sharp step of current. However, we have seen that using ramping currents that are larger than 700 ps suppresses the formation of DS in favor of droplets.

We, thus, speculate about the possibility of observing DS experimentally. Typically, an experimental setup used for the study of droplets contains a free layer with PMA where the solitonic modes may form, which corresponds to our simulated CoNi layer, and a fixed layer that is used as a spin polarizer for the current, \cite{Mohseni2013, Chung2014, Mohseni2014, Macia2014, Lendinez2015, Chung2015, Akerman2016NatComm}. To create a DS instead of a droplet we need to either depart from an initial magnetization state close to all perpendicular or produce torques associated with the Oersted fields larger than those associated with the polarized currents. In the first case we can try to apply large out-of-plane fields in order to set and appropriate initial magnetization state or lower the temperature to reduce the thermal noise that might produce fluctuations of the magnetization. It could be that experiments performed at low temperatures \cite{Macia2014, Lendinez2017} have already created DS. The second case consists in providing a large current that is not polarized, producing large Oersted fields but no STT effect. With the same configuration, the current polarization has to increase so that the STT becomes predominant and promotes the creation of a solitonic mode. If the magnetization was already curled due to the Oersted fields it could result in the creation of a solitonic mode with topological protection: a DS. This realization is feasible by using in-plane polarizers that provide a spin polarization that depends on the out-of-plane applied field. We added in the Supplementary materials simulations where the polarization is varied at a fixed current valued and found that again to create a DS one needs to vary the polarization of the current fast enough---as shown for the applied current, we need pulses of less than 1 ns.

It is necessary however to distinguish experimentally the two solitonic modes once they are created. The differences in precession frequency are two small to serve as a signature of droplet or DS. Instead, studying the stability of the solitonic modes is the best option. One could study the hysteretic response or the response to small in-plane fields and the appearance of low frequency noise as seen in Fig.\ \ref{fig4}.

In conclusion we have shown that both droplet and DS can be created with a same configuration of applied field and spin-polarized current by controlling the initial magnetization state, the degree of spin polarized current, or the speed at which the current--or the polarization---is changed. We also studied the difference in stability between droplet states and DS and found that DS is not only more stable against effective field variations but DS also requires much lower currents to be sustained. Our results provide a pathway for experimental studies of DS and their stability.

\begin{acknowledgments}
F.M. acknowledges financial support from the Ram\'on y Cajal program through RYC-2014-16515 and from MINECO through the Severo Ochoa Program for Centers of Excellence in R\&D (SEV-2015-0496). JMH and NS acknowledge support from MINECO through MAT2015-69144. Research at NYU was supported by Grant No. NSF-DMR-1610416.
\end{acknowledgments}

\newpage
\section*{Supplementary Material for: Generation and stability of dynamical skyrmions and droplet solitons}

\section{Video Description}
The video shows the time evolution of droplet (left-hand-side panel), and a DS (left-hand-side panel) in the presence of an in-plane magnetic field. Both droplet and DS states are created at $30$ mA using different initial states and after stabilization the applied current is reduced to $10$ mA, and a small in-plane field of $50$ mT is applied. Droplet and DS behave completely differently; the droplet has a drift resonance, being annihilated and created again at a frequency of $\sim$ 40 MHz while the DS, although it initially oscillates,  it stabilizes after $\sim80$ ns and remains with the initial topology having $S=1$.

\newpage
\section{Additional simulations}

We calculated a phase diagram of droplet and DS formation as a function the polarization of the applied current and the initial magnetization angle ($\theta_I$) for a fixed current of $30$ mA. The Oersted-field effects are fixed---given by the current of 30 mA---and only STT effects vary with spin-polarization. At small polarizations, there is a small STT effect and no excitations are present independent of the initial magnetization. As the current polarization increases we found first the onset of droplet states and with a further increase the onset of DS. 

\begin{figure}[h]
	 \centering
  \includegraphics[width=0.70\columnwidth]{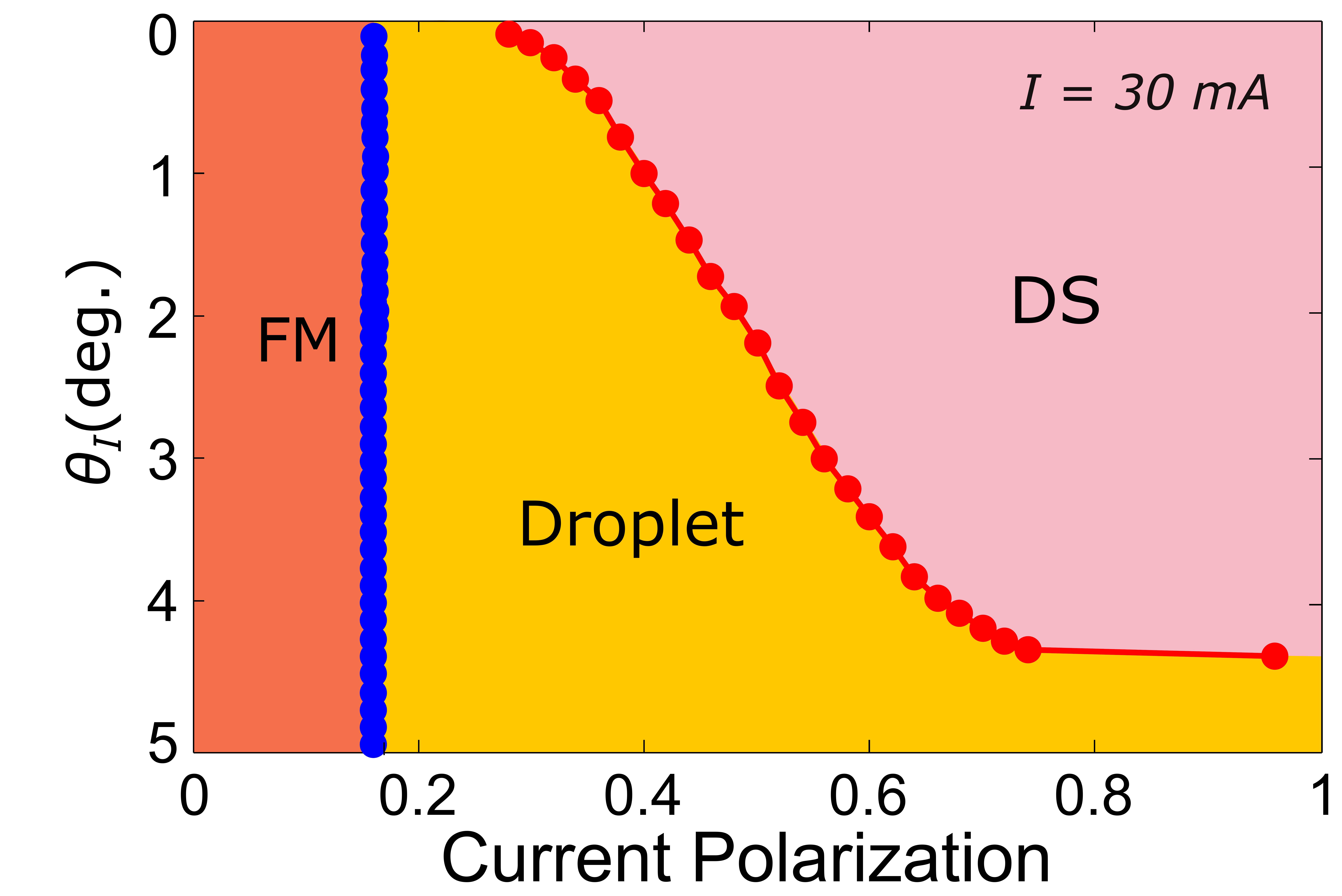}
     \caption{ \textbf{Phase diagram of the droplet and DS formation.} Creation of both solitonic modes as a function of the polarization of the applied current and the initial magnetization angle ($\theta_I$) for a fixed current of $30$ mA. As the current is fixed the Oersted field effects are fixed and only spin-transfer torque (STT) effects vary with spin-polarization.}
     \label{supp2}
\end{figure}

\begin{figure}[h]
	\centering
	\includegraphics[width=0.99\columnwidth]{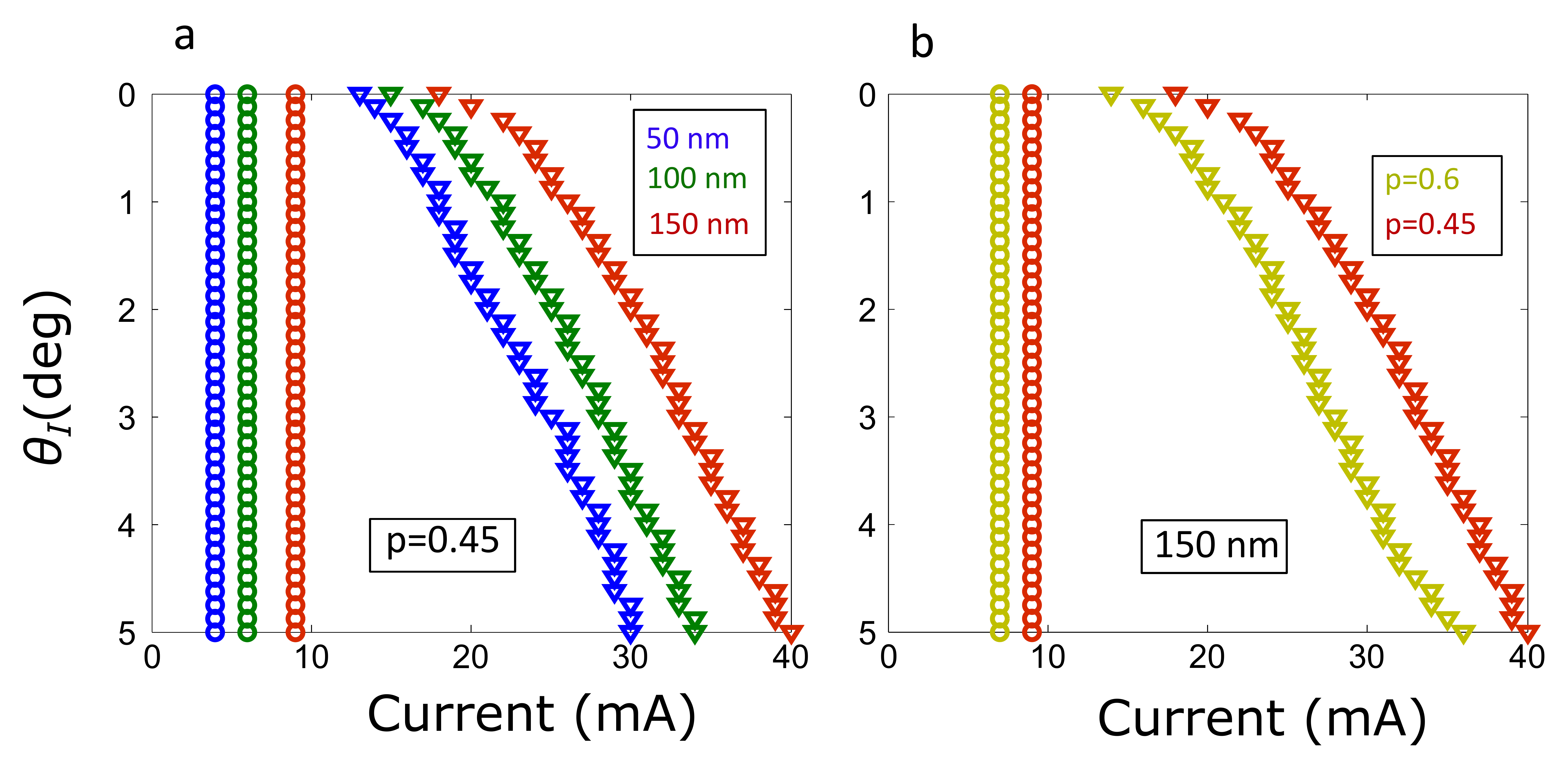}
	\caption{\textbf{Phase diagrams of the droplet and DS formation as a function of NC diameter and polarization.} a) Creation of droplet and DS as a function of applied current for nanocontact diameters of 50, 100, and 150 nm for a fixed current polarization of 0.45. b) Creation of droplet and DS as a function of applied current for current polarizations of 0.45 and 0.6 at a fixed nanocontact diameter of 150 nm.}
	\label{supp3}
\end{figure}

Next we calculated phase diagrams of droplet and DS formation similar to those presented in Fig.\ 3 in the main manuscript for different contact sizes and different current polarizations. In Fig.\ \ref{supp3}a we plot diagrams for different contact size. The contact size determines the total amount of current required to excited solitonic modes. We observe that the total amount of current required for droplet nucleation in contact diameters of 50, 100 and 150 nm is  4, 6, and 9 mA. The threshold for DS creation is also reduced with decreasing the size of the nanocontact with approximately 5 and 9 mA for 50 and 100 nm in comparison with 150nm.
Figure\ \ref{supp3}b compares the same diagram for the 150 nm nanocontact at different current polarization values. An increase of polarization from $p=0.45$ to $p=0.6$ reduced by 2 mA the droplet threshold and about of 5 mA in the DS threshold.

\begin{figure}[h]
	\centering
	\includegraphics[width=0.7\columnwidth]{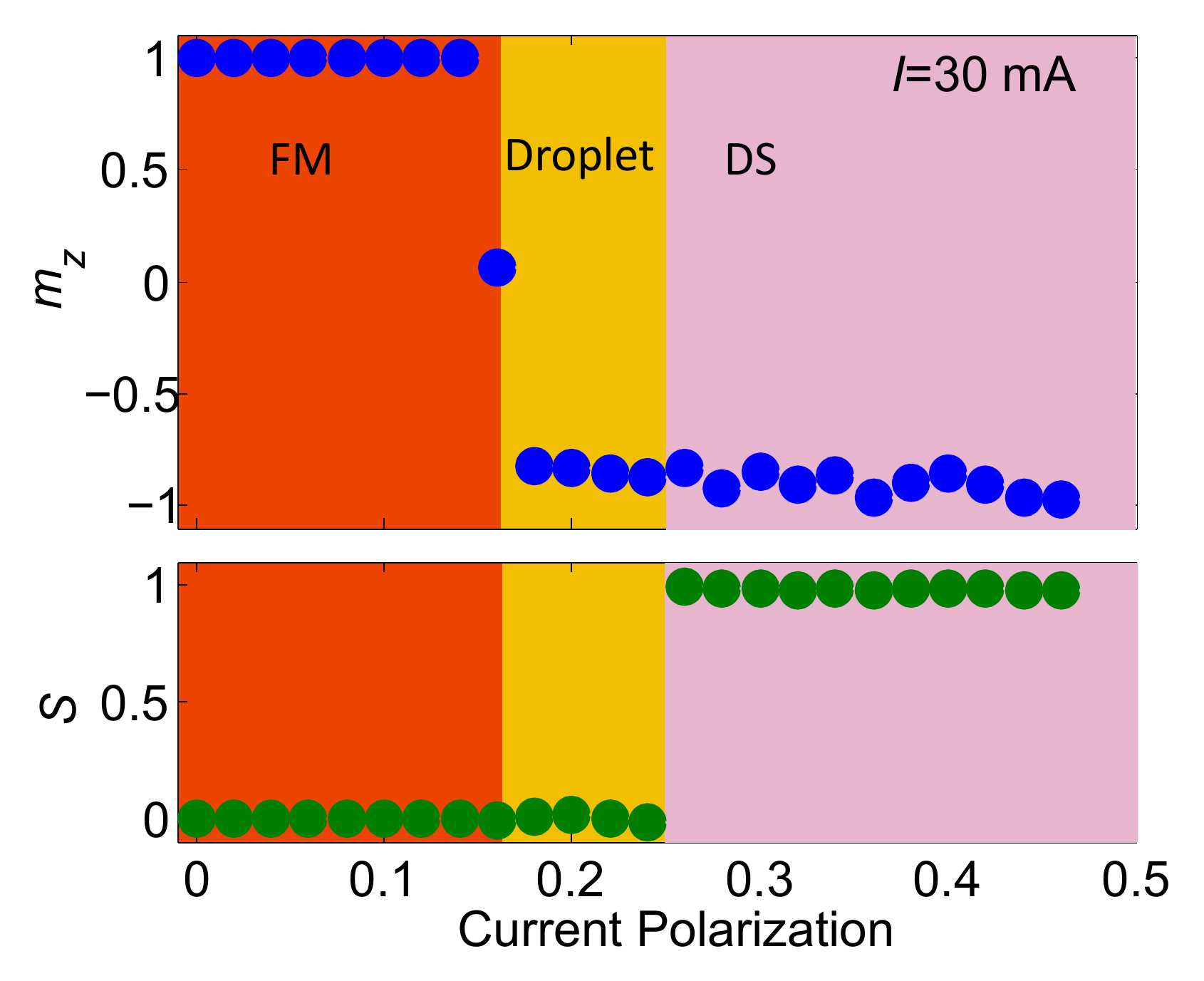}
	\caption{  \textbf{Droplet and dynamical skyrmion creation process}. The upper panel shows the averaged nanocontact magnetization as a function of the current polarization and the lower panel shows the corresponding skyrmion number, $S>$, of the magnetization configuration. A constant current of 30 mA is applied from the beginning for a few ns to allow magnetization relax before a change in polarization is applied. As the polarization increases we cross a first threshold at 0.16 that corresponds to the creation of a droplet. A further increase of polarization creates a DS (above 0.26).}
	\label{supp4}
\end{figure}

Finally we show in Fig.\ \ref{supp4} a simulation where we depart from an equilibrium configuration consisting of having an applied current of 30 mA with no polarization and then changing the polarization to a given value. At low polarizations (below  0.16) no solitonic excitation occurs. The first threshold is at 0.16 and corresponds to a droplet formation whereas the DS formation requires 0.26. The results are similar to what we obtained with increasing current with a fixed polarization but here we always depart from an initial equilibrium configuration. This realization is feasible by using in-plane polarizers that provide a spin polarization that depends on the out-of-plane applied field and varying the out of the plane field to vary the current poalrization.

\newpage
\newpage

\section{Supplementary Note 2: Micromagnetic Code}

\begin{spacing}{0.8}	
	{\footnotesize
			\noindent
			// mumax3 is a GPU-accelerated micromagnetic simulation open-source software\\
			// developed at the DyNaMat group of Prof. Van Waeyenberge at Ghent University.\\
			// The mumax3 code is written and maintained by Arne Vansteenkiste.\\
			\\
			//GRID\\
			NumCells := 256\\
			CellSize:=4.e-9\\
			SetGridSize(NumCells, NumCells, 1)\\
			SetCellSize(CellSize, CellSize, CellSize)\\
			SETPBC(4, 4, 0)			\\
			\\
			//REGIONS\\
			setGeom(layer(0))\\
			diam$_{\text{circ}}$ := 150e-9\\
			r$_{\text{circ}}$ := diam$_{\text{circ}}$ / 2\\
			A$_{\text{circ}}$ := pi * pow(r$_{\text{circ}}$, 2)\\
			DefRegion(1, layer(0).intersect(circle(diam$_{\text{circ}}$)))\\
			\\
			//MATERIAL PARAMETERS FOR STANDARD CoNi\\
			lambda = 1\\
			epsilonprime = 0\\
			Msat = 500e3\\
			Ku1 = 200e3\\
			Aex = 10e-12\\
			alpha = 0.03\\
			anisU = vector(0, 0, 1)\\
			fixedlayer = vector(0., 0., 1.)\\
			\\
			//OERSTED FIELDS\\
			current := vector(0., 0., 1.)\\
			posX := 0.\\
			posY := 0.\\
			mask := newSlice(3, NumCells, NumCells, 1)\\
			\\		
			for i := 0; i $<$ NumCells; i++ \{\\
			for j := 0; j $<$ NumCells; j++ \{\\
					r := index2coord(i, j, 0)\\
					r = r.sub(vector(posX, posY, 0))\\
					b := vector(0, 0, 0)\\
					if r.len() $>=$ r$_{\text{circ}}$ \{\\
						b = r.cross(current).mul(mu0 / (2 * pi * r.len() * r.len()))\\
					\} else \{\\
						b = r.cross(current).mul(mu0 / (2 * pi * r$_{\text{circ}}$ * r$_{\text{circ}}$))\\
					\}\\
					for k := 0; k < 1; k++ \{\\
						mask.set(0, i, j, k, b.X())\\
						mask.set(1, i, j, k, b.Y())\\
						mask.set(2, i, j, k, b.Z())\\
					\}\\
				\}\\
			\}\\
			//RUNNING\\
			B$_{\text{ext}}$.RemoveExtraTerms()\\
			Curr := -30e-3\\
			B$_{\text{ext}}$ = vector(0, 0, 0.5)\\
			Pol = 0.45\\
			Angle := 89.9\\
			my := cos(angle * pi / 180)\\
			mz := sin(angle * pi / 180)\\
			m = Uniform(0, my, mz)\\
			\\
			j.SetRegion(1, vector(0, 0, Curr/A$_{\text{circ}}$))\\
			B$_{\text{ext}}$.RemoveExtraTerms()\\
			B$_{\text{ext}}$.add(mask, Curr)\\
			Run(20e-9)\\
		}
	
	\end{spacing}

\newpage
\bibliography{bibFer}

\end{document}